\documentstyle[12pt]{article}
\begin{document}

\def\a{\alpha}
\def\b{\beta}
\def\ch{\chi}
\def\d{\delta}
\def\e{\epsilon}
\def\f{\phi}
\def\g{\gamma}
\def\h{\eta}
\def\i{\iota}
\def\j{\psi}
\def\k{\kappa}
\def\l{\lambda}
\def\m{\mu}
\def\n{\nu}
\def\o{\omega}
\def\p{\pi}
\def\q{\theta}
\def\r{\rho}
\def\s{\sigma}
\def\t{\tau}
\def\u{\upsilon}
\def\x{\xi}
\def\z{\zeta}
\def\D{\Delta}
\def\F{\Phi}
\def\G{\Gamma}
\def\J{\Psi}
\def\L{\Lambda}
\def\O{\Omega}
\def\P{\Pi}
\def\S{\Sigma}
\def\U{\Upsilon}
\def\X{\Xi}
\def\T{\Theta}

\def\Ab{\bar{A}}
\def\gi{g^{-1}}
\def\li{{ 1 \over \l } }
\def\lb{\l^{*}}
\def\zb{\bar{z}}
\def\ub{u^{*}}
\def\Tb{\bar{T}}
 \def\pp {\partial }
\def\pb {\bar{\partial }}
\def\be{\begin{equation}}
\def\ee{\end{equation}}
\def\ben{\begin{eqnarray}}
\def\een{\end{eqnarray}}

\addtolength{\topmargin}{-0.8in}
\addtolength{\textheight}{1in}
\hsize=16.5truecm
\hoffset=-.5in
\baselineskip=7mm

\thispagestyle{empty}
\begin{flushright} \ March \ 1996\\
SNUTP 96-017 \\
\end{flushright}

\begin{center}
 {\large\bf
 More on Generalized Heisenberg Ferromagnet Models}
\vglue .2in
Phillial Oh\footnote{ E-mail address; ploh@newton.skku.ac.kr }
\vglue .2in
{\it
Department of Physics,
Sung Kyun Kwan University \\
Suwon, 440-746,  Korea}
\vglue .2in
Q-Han Park\footnote{ E-mail address; qpark@nms.kyunghee.ac.kr }
\vglue .2in
{\it Department of Physics, Kyunghee University\\
Seoul, 130-701, Korea}
\vglue .2in
{\bf ABSTRACT}
\end{center}
\vglue .2in

We generalize the integrable Heisenberg ferromagnet model according
to each Hermitian symmetric spaces and address various new aspects
of the generalized model. Using the first order formalism of
generalized spins which are defined on the coadjoint orbits of
arbitrary groups, we construct a Lagrangian of
the generalized model from which we obtain the Hamiltonian structure
explicitly in the case of $CP(N-1)$ orbit.
The gauge equivalence between the generalized Heisenberg ferromagnet and
the nonlinear Schr\"{o}dinger models is given. Using the equivalence, we find
infinitely many conserved integrals of both models.

\newpage
In the past decades, there have been extensive investigations on the
structure of the continuous Heisenberg ferromagnet
(HM) model \cite{fadd}-\cite{tjon}. The dynamical variable of the conventional
HM model is given by a spin variable $Q$ which is defined on  the coadjoint
orbit $S^2$ of $SU(2)$, i.e.
$Q(x,t) = Q^a(x,t)T^a , ~ \sum_{a=1}^{3}Q^a Q^a =k^{2} $,
where $T^a$'s are generators of the $SU(2)$ algebra and $k^{2}$ is a constant.
This $SU(2)$ spin HM model was later extended to the $SU(N)$ case\cite{orfa}.
More generally, it was shown that there exists an extension of the HM model
to each Hermitian symmetric spaces\cite{ford}. However, the extension
in \cite{ford} is made only implicitly in connection with the nonlinear
Schr\"{o}dinger(NS) model. In particular, the integrability structure was not
clear and the Lax pair formalism was lacking.

The purpose of this Letter is to provide a systematic understanding of the
generalized HM model. We formulate the model in terms of a Lagrangian
using the first order formalism of generalized spins which are defined on
the coadjoint orbits of arbitrary groups. The gauge equivalence of the
generalized HM and the generalized NS model is demonstrated. Especially,
starting from the associated linear equation of the HM model, we obtain a
closed form of the generalized NS equation(Eq.({\ref{gnls})).
We also find zero curvature expressions of both the HM and the NS equations
in terms of which we obtain infinitely many conserved integrals.
These conserved integrals are constructed systematically by making use of the
properties of Hermitian symmetric space and they are given in a multicomponent
form thus giving more than ``one series" of integrals.
As an explicit example of the Lagrangian description, we perform a
reduction to $CP(N-1)$ orbit in detail and explain the resulting Hamiltonian
structure of the HM model.

We begin with a brief introduction on the Hermitian symmetric space.
A symmetric space is a coset space $G/K$ for Lie groups $G \supset K$
whose associated Lie algebras ${\bf g}$ and $ {\bf k}$, with the decomposition
${\bf g} = {\bf k} \oplus {\bf m}$, satisfy the commutation relations,
\begin{equation}
[{\bf k} , ~ {\bf k}] \subset {\bf k}, ~~ [{\bf k}, ~ {\bf m}] \subset
{\bf m}, ~~ [{\bf m},~ {\bf m}] \subset {\bf k} .
\label{algebra}
\end{equation}
A Hermitian symmetric space is a symmetric space equipped with a complex
structure. For our purpose, we need only the following properties of
Hermitian symmetric spaces \cite{ford}\cite{helg} ; for each Hermitian symmetric
space, there exists an element $T$ in the Cartan subalgebra of ${\bf g}$
whose centralizer in ${\bf g}$ is ${\bf k}$, i.e. ${\bf k} = \{ V \in
{\bf g}:~ [V,~ T] = 0 \}$. Also, up to a scaling, $J = \mbox{ad}T = [T, ~ *]$
is a linear map $J: {\bf m} \rightarrow {\bf m}$ satisfying the complex structure
condition $J^{2} = \alpha $ for a constant $\alpha $, or
$[T, ~ [T,~ M]] = \alpha M, ~$ for $ M \in {\bf m}$.

Now, we introduce an action for the HM model,
\begin{equation}
A= \int dt dx\mbox{ Tr }[2Tg^{-1}\dot g+
\partial_x (gT g^{-1})\partial_x (gT g^{-1})-2B gTg^{-1}]
\label{actionn}
\end{equation}
where $g$ is a map $g:R^{2} \rightarrow G$ and $B=B(t)$ is an arbitrary
element in ${\bf g}$ describing an external magnetic field.
The equation of motion can be written in terms of a
generalized spin $Q\equiv gTg^{-1}$,
\begin{equation}
\dot Q+\partial_x [Q,~\partial Q]+[Q,~B]=0.
\label{hequation}
\end{equation}
The integrability of the HM equation (\ref{hequation}) arises from the
existence of its associated linear equations:
\be
(\bar{\partial}  -B - \lambda [Q,~\partial Q] + \alpha \lambda^2 Q )
\Psi_{HM} = 0 ,~~~
(\partial   + \lambda  Q)\Psi_{HM} =0,
\label{linear}
\ee
where $\pp = \pp / \pp x , ~ \pb = \pp / \pp t $ and $\lambda $ is an arbitrary
complex constant whereas $\alpha$ is a constant to be fixed later.
These linear equations are overdetermined systems whose
consistency requires the integrability condition:
\ben
0 &=& [\bar{\partial}  -B - \lambda [Q,~\partial   Q] +
\alpha \lambda^2 Q , ~
\partial   + \lambda  Q ] \nonumber \\
&=& \lambda (\bar{\partial} Q+\partial [Q,~\partial Q]+[Q,~B] )
+ \lambda^{2} (- \alpha \partial Q+[Q, ~ [Q,~ \partial  Q]] ).
\label{zero}
\een
The $\l^{1}$-order term in the second line of Eq.(\ref{zero}) becomes
precisely the HM equation since the $\l^{2}$-order term vanishes
identically due to the complex structure property of $T$,
\ben
[Q, ~ [Q,~ \partial  Q]] &=& g[T, ~ [T, ~ g^{-1}(\partial  Q)g]]g^{-1}
= g[T, ~ [T, ~ [g^{-1}\partial  g, ~ T]]]g^{-1} \nonumber \\
&=& \alpha g[g^{-1}\partial  g, ~ T]g^{-1} = \alpha \partial  Q ,
\een
or, whenever $T$ satisfies a stronger condition:
\begin{equation}
T^2=\beta T +\gamma I, ~~ \alpha = \beta^{2} + 4\gamma .
\label{condk}
\end{equation}
Without loss of generality, we may set $\beta $ to zero by shifting
$T$ by $T \rightarrow T - \beta /2$. This stronger condition holds
at least for $SU(p+q)/S(U(p) \times U(q)), ~ SO(2n)/U(n), ~ Sp(n)/U(n)$
compact Hermitian symmetric spaces and their noncompact counterparts.
Since $K$ is a subgroup  of $G$ commuting with $T$, this shows that
$Q$ in the case of Eq.(\ref{condk}) is in fact defined on the coadjoint
orbit $G/K$ characterized by a number $ Q^{2} = \g I$.
Note that the external magnetic field $B(t)$
can be made disappear by taking the gauge transformation of $g$ and $\Psi_{HM} $
such that $g \rightarrow b^{-1}(t)g, ~~ \Psi_{HM} \rightarrow b^{-1}(t)\Psi_{HM} $
where $b(t)$ satisfies $\bar{\partial} bb^{-1} = B(t)$. From now on, we
assume that $B=0$.

Having found the associated linear equations of the HM equation, we demonstrate
the integrability of the HM model itself by deriving infinitely many
conserved currents from the linear equations. In order to do so, we first
derive the generalized NS equation from the generalized HM equation thereby
proving the gauge equivalence of both models.  Define
$\Psi_{NS} \equiv g^{-1} \Psi_{HM} $ and rewrite the linear equation (\ref{linear})
in an equivalent form:
\be
(\bar{\partial}  + g^{-1}\bar{\partial}  g - \lambda g^{-1}\partial   g
-\lambda^{2} T )\Psi_{NS} = 0 ,~~~
(\partial   + g^{-1}\partial   g+ \lambda  T)\Psi_{NS} =0 ,
\label{lin2}
\ee
where we have taken $\alpha = -1$ without loss of generality. Since $Q$ is
invariant under $g \rightarrow gk $ for $k \in K$, we choose $k$ such that
$g^{-1}\pp g$ is valued in ${\bf m}$.
The integrability of Eq.(\ref{lin2}) becomes the zero curvature condition:
\be
 [\bar{\partial}  + g^{-1}\bar{\partial}  g - \lambda g^{-1}\partial   g
-\lambda^{2} T , ~
\partial   + g^{-1}\partial   g+ \lambda  T ]  = 0 .
\ee
Equivalently, we may require
\be
[T, ~ g^{-1}\bar{\partial}  g] - \partial   ( g^{-1}\partial   g ) = 0
\label{nls1}
\ee
and the identity
\be
[\bar{\partial}  + g^{-1}\bar{\partial}  g, ~ \partial   + g^{-1}\partial   g]
=0.
\label{nls2}
\ee
$g^{-1} \bar{\partial}  g $ may be expressed in terms of $g^{-1} \partial   g $
by solving Eqs.(\ref{nls1}) and (\ref{nls2}) as follows;
introduce a decomposition
 $ g^{-1} \bar{\partial}  g  =  (g^{-1} \bar{\partial}  g )_{m} +
(g^{-1} \bar{\partial}  g )_{k}$ where subscript $m$ and $k$ refer to the
components of  $g^{-1} \bar{\partial}  g $ in those vector subspaces.
Then, due to the algebraic properties of Eq.(\ref{algebra}), Eq.(\ref{nls1})
becomes
\be
 [T , ~ (g^{-1} \bar{\partial}  g)_{m} ] - \partial   ( g^{-1}\partial   g ) = 0
\ee
which can be solved for $(g^{-1} \bar{\partial}  g)_{m}$ by applying the adjoint
action of $T$,
\be
[T, ~ [T , ~ (g^{-1} \bar{\partial}  g)_{m} ] = - (g^{-1} \bar{\partial}  g)_{m}
=  [T , ~ \partial   (g^{-1} \partial   g) ] .
\label{int1}
\ee
The $k$-component of $g^{-1} \bar{\partial}  g$ can be obtained from Eq.(\ref{nls2})
which decomposes into
\be
0 = \partial   (g^{-1} \bar{\partial}  g)_{m} - \bar{\partial}  (g^{-1} \partial
g ) + [ g^{-1} \partial   g , ~ (g^{-1} \bar{\partial}  g)_{k} ]
\label{nls4}
\ee
and
\ben
0 &=& \partial   (g^{-1} \bar{\partial}  g)_{k} + [ g^{-1} \partial   g, ~
(g^{-1} \bar{\partial}  g)_{m} ]  \nonumber \\
& = & \partial   (g^{-1} \bar{\partial}  g)_{k} -
[ g^{-1} \partial   g, ~ [T, ~ \partial   (g^{-1} \partial   g)]] \nonumber \\
&=& \partial   (g^{-1} \bar{\partial}  g)_{k} +
\frac{1}{2}\partial   [ g^{-1} \partial   g, ~
[ g^{-1} \partial   g, ~T]].
\label{nls3}
\een
This may be integrated directly to yield
\be
(g^{-1} \bar{\partial}  g)_{k} =
{1\over 2} [g^{-1} \partial   g, ~ [T, ~  g^{-1} \partial   g ]] +C(t)
\label{int2}
\ee
where the arbitrary function $C(t)$ can be set to zero by
redefining $g$. Finally, the remaining equation (\ref{nls4}) becomes the
\underline{generalized NS equation}:
\be
\bar{\partial}  ( g^{-1} \partial   g ) + [T, ~ \partial  ^{2}
( g^{-1} \partial   g )] + {1\over 2}[ g^{-1} \partial   g , ~
[  g^{-1} \partial   g , ~ [  g^{-1} \partial   g, ~ T]]] =0.
\label{gnls}
\ee
Thus we have shown the gauge equivalence of the HM equation and the NS equation.

Such a gauge equivalence also relates the conserved integrals of both models.
In order to find the conserved integrals of the NS model, we first rewrite
the linear equation in an equivalent form with $\Phi =
 \Psi_{NS}\exp(\lambda Tx - \lambda^{2}Tt) $,
\be
\bar{\partial}  \Phi + (g^{-1} \bar{\partial}  g - \lambda g^{-1}
\partial   g)\Phi -\lambda^{2} [T, ~ \Phi ]=0, ~~
\partial   \Phi + g^{-1} \partial   g \Phi + \lambda[T, ~ \Phi ] = 0 ,
\label{cons1}
\ee
which we solve iteratively by assuming
\be
\Phi = \sum_{l=0}^{\infty }{1 \over \lambda^{l} }(\Phi^{l}_{m} +
\Phi^{l}_{k}).
\label{itera}
\ee
The subscript denotes the decomposition of $\Phi$ with the properties
\be
[T, ~ \Phi_{k}] = 0, ~ \Phi_{k}\Phi_{k} \subset \Phi_{k} , ~
\Phi_{k}\Phi_{m} \subset \Phi_{m} , ~ \Phi_{m}\Phi_{m} \subset \Phi_{k} .
\label{decomrel}
\ee
Explicit construction of such a matrix decomposition can be carried out
for each Hermitian symmetric spaces given in \cite{ford}. For example, in 
the case of $SU(p+q)/S(U(p) \times U{q})$, we note that any $U(p+q)$ group
element $\Phi $ can be expressed as a sum of $\Phi_{m} = \sum c_{i}M^{i} $
and $\Phi_{k} = d_{0}I + \sum d_{j}K^{j}$, where $I$ is the identity matrix,
$M^{i}$'s are basis vectors for the symmetric space
$SU(p+q)/S(U(p) \times U(q))$ and $K^{j}$'s are the generators
of the subalgebra ${\bf k}$. Obviously, this decomposition satisfies the
relation Eq.(\ref{decomrel}). With Eq.(\ref{decomrel}),
Eq.(\ref{cons1}) changes into a set of recursive relations,
\ben
\partial   \Phi^{l}_{m} +  g^{-1} \partial   g \Phi^{l}_{k}
 &=& - [T, ~ \Phi^{l+1}_{m}]
\label{recu1}
\\
\partial   \Phi^{l}_{k} + g^{-1} \partial   g \Phi^{l}_{m} &=& 0
\label{recu2}
\een
and
\ben
\bar{\partial}  \Phi^{l}_{m} + (g^{-1} \bar{\partial}  g )_{m} \Phi^{l}_{k} +
(g^{-1} \bar{\partial}  g )_{k} \Phi^{l}_{m}- g^{-1} \partial   g
\Phi^{l+1}_{k} &=& [T, ~ \Phi^{l+2}_{m}]
\label{recu3}
 \\
\bar{\partial}  \Phi^{l}_{k} + (g^{-1} \bar{\partial}  g )_{m} \Phi^{l}_{m} +
(g^{-1} \bar{\partial}  g )_{k} \Phi^{l}_{k}- g^{-1} \partial   g
\Phi^{l+1}_{m} &=& 0 .
\label{recu4}
\een
Eq.(\ref{recu4}) may be rewritten by using Eqs.(\ref{int1}), (\ref{int2})
and (\ref{recu1}),
\be
\bar{\partial}  \Phi^{l}_{k} -[T, ~ \partial  (g^{-1} \partial   g )]
\Phi^{l}_{m} - g^{-1} \partial   g [T, ~ \partial   \Phi^{l}_{m} ] = 0.
\label{recu5}
\ee
Eqs.(\ref{recu1}),(\ref{recu2}) and (\ref{recu5}) may be solved for
$\Phi^{l}$ in lower order terms of iteration,
\ben
\Phi^{l}_{m} &=& [T, ~ \partial   \Phi^{l-1}_{m} +  g^{-1} \partial   g
\Phi^{l-1}_{k} ] \nonumber \\
\Phi^{l}_{k} &=& -\int dx  g^{-1} \partial   g \Phi^{l}_{m} +
\int dt ( [T, ~ \partial  (g^{-1} \partial   g )] \Phi^{l}_{m}
+ g^{-1} \partial   g [T, ~ \partial   \Phi^{l}_{m} ]) .
\label{itersol}
\een
Then, from the compatibility of Eqs.(\ref{recu2}) and (\ref{recu5})
($\partial   \bar{\partial}  \Phi^{l}_{k}
= \bar\partial   {\partial}  \Phi^{l}_{k} $ ), we obtain infinitely
many conserved currents,
\be
\bar{\partial}  J^{l}_{x} + \partial  J^{l}_{t} = 0 ~ ; ~~ l = 0,1,2, \cdots
\ee
where
\ben
J_{x}^{l} &=& - \pp \Phi_{k}^{l} =  g^{-1} \partial   g \Phi^{l}_{m} \nonumber \\
J_{t}^{l} &=& \pb \Phi_{k}^{l} =  [T, ~ \partial  (g^{-1} \partial   g )] \Phi^{l}_{m}
+ g^{-1} \partial   g [T, ~ \partial   \Phi^{l}_{m} ] .
\een
In particular, if we choose an initial condition $\Phi^{0}_{m} = 0, ~
\Phi^{0}_{k} = 1$, we obtain for $l=1$,
\be
J_{x}^{1} = g^{-1} \partial   g [T, ~ g^{-1} \partial   g] , ~~
J_{t}^{1} = [T, ~ \partial  (g^{-1} \partial   g )]
[T, ~ g^{-1} \partial   g ] - g^{-1} \partial   g
\partial  (g^{-1} \partial   g ) .
\label{nscon}
\ee
In the simplest case where $G/K = SU(2)/U(1)$ or $ SU(1,1)/U(1) $, we may take $g^{-1} \pp g$ and $T$ by
\be
g^{-1}\pp g = \pmatrix{ 0 & \kappa \psi \cr
			\kappa \psi^{*} & 0 } , ~~~~
T = \pmatrix{ {i \over 2} & 0 \cr
	0 & -{i \over 2}}
\ee
where $\kappa = 1 $  for the compact $SU(2)/U(1)$ case and $\kappa = i$ for the
noncompact $SU(1,1)/U(1)$ case.
Then Eq.(\ref{gnls}) reduces to the conventional nonlinear Schr\"{o}dinger
equation
\be
\pb \psi + i\pp^{2} \psi - 2i\kappa^{2}|\psi |^{2}\psi = 0 .
\ee
In particular, the $T$-component of the conserved current in
Eq.(\ref{nscon}), i.e
\ben
 0 &=& \mbox{ Tr }T(\pb J_{x}^{1} + \pp J_{t}^{1})
\nonumber \\
&=& \pb ( \psi \psi^{*} ) + \pp (i\psi^* \pp \psi - i\psi \pp \psi^* )
\een
corresponds to the probability conservation.
Note that our conservation laws are given in a matrix form
thus resulting more than one series of conserved local integrals.
This is consistent with our generalized NS model as a multicomponent
system.

The conserved integrals of the HM model can be derived from those of
the NS model by using the gauge equivalence.
In order to do so, we rewrite the iterative solution (\ref{itersol})
in terms of $\Sigma_{m}^{l} \equiv g\Phi_{m}^{l} g^{-1}$ and
$\Sigma_{k}^{l} \equiv g\Phi_{k}^{l} g^{-1}$ such that
\ben
\Sigma_{m}^{l} &=& [ Q, ~ \partial \Sigma_{m}^{l-1} - [[Q,~ \partial Q]
,\Sigma_{m}^{l-1}]]-\partial Q\Sigma_{k}^{l-1} \nonumber \\
\Sigma_{k}^{l} &=& - \int dx J_{x}^{l} + \int dt J_{t}^{l} ,
\een
where the conserved currents $J_{x}^{l}, ~ J_{t}^{l}$  are given by
\ben
J_{x}^{l}  &\equiv & - \partial \Sigma_{k}^{l} =  -[[Q, \partial Q], ~
\Sigma_{k}^{l} ] + [Q, \partial Q]\Sigma_{m}^{l}  \nonumber \\
J_{t}^{l} &\equiv &\bar \partial \Sigma_{k}^{l} = [\partial^{2} Q +
{1\over 2}[\partial Q , ~ [Q, ~ \partial Q ]], ~ \Sigma_{k}^{l}]
- (\partial^{2} Q+ [\partial Q , ~ [Q, ~ \partial Q ]])
\Sigma _{m}^{l} \nonumber
\\
&& + [Q, \partial Q][Q, ~ \partial \Sigma_{m}^{l} -  [[Q,
\partial Q], ~ \Sigma_{m}^{l}]].
\een
Obviously, these integrals, even though there are infinitely many, do not
exhaust all conserved integrals of the HM model. For instance,
the HM equation itself represents a conservation law
$\pb Q + \pp [Q, ~ \pp Q] = 0$ but this does not arise from the above
construction. This lacking is due to the specific choice of iterative solutions as
in Eq.(\ref{itera}) and can be avoided by taking a different kind of
iterative solutions.

For the rest of the Letter, we assume the stronger condition (\ref{condk})
and present, as an example of Lagrangian description of the HM model,
an explicit reduction to the $CP(N-1)$ orbit.
Let us first consider $T$  given by
\begin{equation}
T=\mbox{diag}(t_1,\cdots, t_N),\quad
t_1=\cdots =t_{n}=a, ~~
t_{n+1}=\cdots =t_N=b
\label{firsts}
\end{equation}
with the traceless condition $ na+(N-n)b=0$.  $T$ satisfies
the stronger condition (\ref{condk})
with $\beta=[(2n-N)/(n-N)]a, \ \gamma=a^2n/(N-n)$.
The isotropy group $K$ for this element $T$ is given by
$K=SU(n)\times SU(N-n)\times U(1)$ and the coadjoint orbit
is the Grassmannian manifold $SU(N)/(SU(n)\times SU(N-n)\times U(1))$.
For the non-compact case, it corresponds to
$SU(n,N-n)/(SU(n)\times SU(N-n)\times U(1))$.
For other cases of Hermitian symmetric spaces,
we may take an $2n \times 2n $
matrix $T$ by
\begin{equation}
T=\left(\begin{array}{cccccc}
   0 & -t & \cdot&\cdot &\cdot& 0 \\
  t & 0 & \cdot&\cdot &\cdot& 0 \\
    \vdots &   & \ddots &    &   & \\
    0 &  & & &0 & -t \\
    0 &  & & & t& 0
 \end{array} \right).
\label{elem}
\end{equation}
so that $\beta=0,\gamma=-t^2$. Then, the isotropy group for the above  $T$
is $U(n) $ and the corresponding integrable coadjoint orbits
are given by  \cite{helg} $SO(2n)/U(n)$, $Sp(2n)/U(n)$
and their noncompact counterparts.
The remaining Hermitian symmetric spaces \cite{ford,helg}
do not satisfy the stronger condition.

Now, we restrict to the $CP(N-1) = SU(N)/(SU(N-1) \times U(1))$ case.
In this case, $T=i \mbox{diag}(N-1,-1,\cdots, -1)$
 and let us parametrize the group element $g$ of $SU(N)$
by an $N$-tuple $ g = (Z_1, Z_2,\cdots, Z_N),
\ Z_p\in {\bf C}^N$ $(p,q=1,\cdots,N)$ such that
 \begin{equation}
\bar Z_pZ_q=\delta_{pq}, \quad \mbox{det}
(Z_1, Z_2,\cdots, Z_N) =1.
\label{cond}
\end{equation}
With this parametrization, the action (\ref{actionn}) can be written
in terms of $Z$,
\begin{equation}
A=\int dt dx  \left[2Ni\bar Z_1 \dot Z_1-2N^2
(\partial \bar Z_1\partial Z_1-
(\bar Z_1\partial  Z_1)(\partial\bar Z_1 Z_1))+\lambda (\bar Z_1 Z_1-1)
\right]\label{action2}
\end{equation}
where we imposed the constraint of Eq.(\ref{cond}) by introducing a
Lagrangian multiplier $\l $.
Note that all the $Z_p$'s with $p=2,\cdots,N$
have been eliminated and they do not appear in the subsequent
analysis.
The above action is invariant under the local $U(1)$ action,
$Z_1\rightarrow e^{i\alpha (x,t)}Z_1,
\bar Z_1\rightarrow e^{-i\alpha (x,t)}\bar Z_1$,
and the constraint $\bar Z_1 Z_1-1\approx 0$ is  a first class constraint.
Let $Z_1^T \equiv (z_1,z_2,\cdots,z_N)$. Then, the constraint can be solved
explicitly in a  real gauge where $z_N=z^*_N$ $(z_N\neq 0)$
\cite{alek}  in terms of $\psi_i=z_i/z_N$$(i=1,2,\cdots,N-1)$ so that
\begin{equation}
z_i=\frac{\psi_i}{\sqrt{1+\vert\psi\vert^2}},\quad
z_N=\frac{1}{\sqrt{1+\vert\psi\vert^2}} .
\label{slp}
\end{equation}
Substituting the above expression into the action (\ref{action2}),
we obtain a reduced action on the $CP(N-1)$ orbit (up to a total
derivative term),
\begin{equation}
A=\int dt dx\left(2iN\frac{\bar \psi_i\dot\psi_i}
{1+\vert\psi\vert^2}-2N^2 g_{ij}\partial \psi_i
\partial \bar\psi_j\right),
\end{equation}
where $g_{ij}$ is the Fubini-Study metric on $CP(N-1)$,
\begin{equation}
g_{ij}=\frac{(1+\vert\psi\vert^2)\delta_{ij}
-\bar\psi_i\psi_j}{(1+\vert\psi\vert^2)^2}.
\end{equation}
Note that the first term in the above action can be written as
$2N\int dx \theta$ where $\theta$ is the canonical one-form
on $CP(N-1)$ defined by $\theta=i\partial_{\psi} \log(1+\vert\psi\vert^2)d\psi $.
The classical dynamics of the above action can be described
by a generalized Hamiltonian dynamics \cite{arno,fadd4}
with the Hamiltonian given by
\begin{equation}
H=2N^2\int dx g_{ij}\partial \psi_i
\partial \bar\psi_j.
\end{equation}
The Poisson bracket is defined by the
inverse matrix $\omega^{ij}=-ig^{ij}$
of the symplectic two-form $\omega=d\theta=w_{ij}d\psi_id\bar\psi_j$ such
that
\begin{equation}
\{F(\bar\psi,\psi),G(\bar\psi,\psi)\}=(2iN)^{-1}
\int dx ~g^{ij}\left(\frac{\delta F
}{\delta \bar \psi_i(x)}\frac{\delta G}{\delta\psi_j(x)}-
\frac{\delta G}{\delta \bar\psi_i(x)}
\frac{\delta F}{\delta \psi_j(x)}\right)
\label{pbracket}
\end{equation}
with the inverse Fubini-Study metric $g^{ij}=(1+\vert\psi\vert^2)
(\delta_{ij}+\bar\psi_i\psi_j)$.
A simple calculation gives
\begin{equation}
\{\bar\psi_i(x,t),\psi_j(x^\prime,t)\}=(2iN)^{-1}g^{ij}\delta(x-x^\prime),\ \ \
\{\bar\psi_i(x,t),\bar\psi_j(x^\prime,t)\}=\{\psi_i(x,t),\psi_j(x^\prime,t)\}=0.
\end{equation}

The Hamiltonian description of the above model in terms of a 
generalized $SU(N)$ spin is more efficient because of the $SU(N)$ invariance.
Consider the unreduced $Q$,
\begin{equation}
Q=iNZ_1\bar Z_1-iI.
\label{isosp}
\end{equation}
Defining the generalized spin functions by $Q^a=2\mbox{Tr}(QT^a)$
in which $T^a$'s are generators satisfying the commutation relation
$[T^a,T^b]=f^{abc}T^c$
and the normalization Tr$(T^aT^b)=(-1/2)\delta_{ab}$,
we have
\begin{equation}
Q^a(\psi,\bar\psi)=2iN\sum_{p,q=1}^{N}\bar z_p(T^a)_{pq}z_q
\label{spinfunction}
\end{equation}
where Eq.(\ref{slp}) is assumed.
A straightforward  computation using Eq.(\ref{pbracket}) gives
\begin{equation}
\{Q^a(x,t),Q^b(x^\prime,t)\}=-f^{abc}Q^c(x,t)\delta(x-x^\prime).
\end{equation}
Also, using the Hamiltonian expressed in terms of $Q^a$'s,
$H=(1/2)\int dx (\partial Q^a)^2$, we recover
 the equation of motion (\ref{hequation}) with $B=0$:
\begin{equation}
\dot Q^a=\{H, Q^a\}=f^{abc}Q^b\partial ^2Q^c.
\end{equation}

In this Letter, we have shown that
the HM model and the NS model can be generalized according to each
Hermitian symmetric spaces and their integrability structures
can be studied systematically by making use of the properties of
Hermitian symmetric space. These approach may be
extended to the study of other properties of the generalized model, e.g.
construction of soliton solutions and their scattering behaviors both classical
and quantum.  Also, our Lagrangian of the HM model might be used in
quantizing the model through the path integral approach.
These works are in progress and will be reported elsewhere
\cite{ohp}.

\vglue .2in
{\bf ACKNOWLEDGEMENT}
\vglue .2in
We like to  thank  Prof. H.J. Shin  for useful discussions.
This work is supported in part by the program of Basic Science Research,
Ministry of Education  BSRI-95-2442/BSRI-95-1419 , and
by Korea Science and Engineering Foundation through the Center for
Theoretical Physics, SNU.


\begin{thebibliography}{99}
\bibitem{fadd} L. D. Faddeev and L. A. Takhtajan,
Hamiltonian Methods in the Theory of Solitons (Springer-Verlag,
Berlin, 1987).
\bibitem{laks} M. Lakshmanan, Phys. Lett.  A 61 (1977) 53.
\bibitem{takh} L. A. Takhtajan, Phys. Lett. A 64 (1977) 235.
\bibitem{tjon} J. Tjon and J. Wright, Phys.  Rev.  B 15 (1977) 3470.
\bibitem{orfa} S. J. Orfanidis, Phys. Lett.  A 75 (1980) 304.
\bibitem{ford} A. P. Fordy and P. P. Kulish, Commum. Math. Phys. 89
(1983) 427.
\bibitem{helg} S. Helgason,  Differential geometry, Lie groups and symmetric
spaces 2nd ed. (New York, Academic Press, 1978).
\bibitem{alek} A. Alekseev, L.D. Faddeev and S. L. Shatashvili, in
  Geometry and Physics  eds.  S. Gindikin and I. M. Singer
(North-Holland, Amsterdam, 1991) p.391;
  T. Lee and P. Oh, Phys. Lett. B  319  (1993) 497.
\bibitem{arno} R. Abraham and J. E. Marsden,  Foundations of Mechanics
	  (Addison Wesley, New York, 1978).
\bibitem{fadd4} P. A. M. Dirac,  Lectures on Quantum Mechanics
(Yeshiva Univ., New York, 1964);
 L. D. Faddeev and R. Jackiw, Phys. Rev. Lett.  60 (1988) 1692.
\bibitem{ohp} P. Oh and Q-H. Park, in preparation.
\end{thebibliography}
\end{document}